\documentstyle[12pt]{article}
\input epsf

\ifx\epsffile\undefined
\newlength{\epsfysize}
\def\epsffile#1{}
\else
\fi

\catcode`\@=11
\long\def\@makefntext#1{
\protect\noindent \hbox to 3.2pt {\hskip-.9pt
$^{{\ninerm\@thefnmark}}$\hfil}#1\hfill}		

\def\@makefnmark{\hbox to 0pt{$^{\@thefnmark}$\hss}}  
\def\ps@myheadings{\let\@mkboth\@gobbletwo
\def\@oddhead{\hbox{}
\rightmark\hfil\ninerm\thepage}
\def\@oddfoot{}\def\@evenhead{\ninerm\thepage\hfil
\leftmark\hbox{}}\def\@evenfoot{}
\def\sectionmark##1{}\def\subsectionmark##1{}}

\renewcommand{\thefootnote}{\fnsymbol{footnote}}

\newcounter{sectionc}\newcounter{subsectionc}\newcounter{subsubsectionc}
\renewcommand{\section}[1] {\vspace*{0.6cm}\addtocounter{sectionc}{1}
\setcounter{subsectionc}{0}\setcounter{subsubsectionc}{0}\noindent
	{\normalsize\bf\thesectionc. #1}\par\vspace*{0.4cm}}
\renewcommand{\subsection}[1] {\vspace*{0.6cm}\addtocounter{subsectionc}{1}
	\setcounter{subsubsectionc}{0}\noindent
	{\normalsize\it\thesectionc.\thesubsectionc. #1}\par\vspace*{0.4cm}}
\renewcommand{\subsubsection}[1]
{\vspace*{0.6cm}\addtocounter{subsubsectionc}{1}
	\noindent {\normalsize\rm\thesectionc.\thesubsectionc.\thesubsubsectionc.
	#1}\par\vspace*{0.4cm}}

\newcounter{appendixc}
\newcounter{subappendixc}[appendixc]
\newcounter{subsubappendixc}[subappendixc]

\renewcommand{\appendix}[1] {\vspace*{0.6cm}
        \refstepcounter{appendixc}
        \setcounter{figure}{0}
        \setcounter{table}{0}
        \setcounter{equation}{0}
        \renewcommand{\thefigure}{\Alph{appendixc}.\arabic{figure}}
        \renewcommand{\thetable}{\Alph{appendixc}.\arabic{table}}
        \renewcommand{\theappendixc}{\Alph{appendixc}}
        \renewcommand{\theequation}{\Alph{appendixc}.\arabic{equation}}
        \noindent{\bf Appendix \theappendixc #1}\par\vspace*{0.4cm}}

\def\abstracts#1{{
\centering{\begin{minipage}{12.2truecm}\vspace*{.1cm}
        \footnotesize\baselineskip=12pt\noindent
	\parindent=0pt #1
	\end{minipage}}\par}}

\renewenvironment{thebibliography}[1]
	{\begin{list}{\arabic{enumi}.}
	{\usecounter{enumi}\setlength{\parsep}{0pt}
\setlength{\leftmargin 1.25cm}{\rightmargin 0pt}
	 \setlength{\itemsep}{0pt} \settowidth
	{\labelwidth}{#1.}\sloppy}}{\end{list}}

\newcounter{itemlistc}
\newcounter{romanlistc}
\newcounter{alphlistc}
\newcounter{arabiclistc}

\newcommand{\fcaption}[1]{
        \refstepcounter{figure}
        \setbox\@tempboxa = \hbox{\footnotesize Fig.~\thefigure. #1}
        \ifdim \wd\@tempboxa > 6in
           {\begin{center}
        \parbox{6in}{\footnotesize\baselineskip=12pt Fig.~\thefigure. #1}
            \end{center}}
        \else
             {\begin{center}
             {\footnotesize Fig.~\thefigure. #1}
              \end{center}}
        \fi}

\newcommand{\tcaption}[1]{
        \refstepcounter{table}
        \setbox\@tempboxa = \hbox{\footnotesize Table~\thetable. #1}
        \ifdim \wd\@tempboxa > 6in
           {\begin{center}
        \parbox{6in}{\footnotesize\baselineskip=12pt Table~\thetable. #1}
            \end{center}}
        \else
             {\begin{center}
             {\footnotesize Table~\thetable. #1}
              \end{center}}
        \fi}

\def\@citex[#1]#2{\if@filesw\immediate\write\@auxout
	{\string\citation{#2}}\fi
\def\@citea{}\@cite{\@for\@citeb:=#2\do
	{\@citea\def\@citea{,}\@ifundefined
	{b@\@citeb}{{\bf ?}\@warning
	{Citation `\@citeb' on page \thepage \space undefined}}
	{\csname b@\@citeb\endcsname}}}{#1}}

\newif\if@cghi
\def\cite{\@cghitrue\@ifnextchar [{\@tempswatrue
	\@citex}{\@tempswafalse\@citex[]}}
\def\citelow{\@cghifalse\@ifnextchar [{\@tempswatrue
	\@citex}{\@tempswafalse\@citex[]}}
\def\@cite#1#2{{$\null^{#1}$\if@tempswa\typeout
	{IJCGA warning: optional citation argument
	ignored: `#2'} \fi}}

 1
 1
 1

\font\ninerm=cmr9

\def\be{ \begin{equation} }
\def\ee{ \end{equation} }
\def\bea{\begin{eqnarray}}
\def\eea{\end{eqnarray}}
\def\ie{{\it i.e.}}
\def\half{  {1\over 2} }

\begin{document}
\begin{flushright}
BOW-PH--107 \\
hep-ph/9507350 \\
\end{flushright}
\vspace{0.1cm}

\centerline{\normalsize\bf ELECTROWEAK STRINGS AND FERMIONS}

\vspace*{0.6cm}
\centerline{\footnotesize SHION KONO}
\vspace*{0.1cm}
\centerline{\footnotesize\it and}
\vspace*{0.1cm}
\centerline{\footnotesize STEPHEN G. NACULICH}
\vspace*{0.3cm}
\baselineskip=13pt
\centerline{\footnotesize\it Department of Physics}
\baselineskip=12pt
\centerline{\footnotesize\it Bowdoin College}
\centerline{\footnotesize\it Brunswick, ME  04011, U.S.A.}
\baselineskip=13pt

\vspace*{0.9cm}
\abstracts{Z-strings in the Weinberg-Salam model
including fermions are unstable for all values of the parameters.
The cause of this instability is the fermion vacuum energy
in the Z-string background.  Z-strings with non-zero fermion
densities, however, may still be stable.  }

\def\d { {\rm d} }
\def\e { {\rm e} }
\def\ppm{ \psi_\pm }
\def\P{ P_\pm}
\def\Ee{E_{\rm effective}}
\def\Eb{E_{\rm boson}}
\def\Ef{E_{\rm fermion}}
\def\Efz{E_{{\rm fermion}}^\prime}
\def\Cb{C_{\rm b}}
\def\Cf{C_{\rm f}}
\def\eps{\epsilon}
\def\del{\partial}
\def\Dslash{\rlap{D}\,/}
\def\Phitil{{\tilde \Phi}}
\def\Psibar{{\overline \Psi}}
\def\psibar{{\overline \psi}}
\def\rrho{\rho^\prime}

\normalsize\baselineskip=15pt
\setcounter{footnote}{0}
\renewcommand{\thefootnote}{\alph{footnote}}

\section{Electroweak strings}

Over the last two decades,
extended objects in field theories called solitons
have played an important role in particle physics and astrophysics.
In certain cases,
these solitons possess a conserved topological charge,
which guarantees their stability.
There are other extended field configurations, however,
which are not topological;
these ``nontopological solitons'' are stable for dynamical reasons:
they sit at a local minimum of the energy functional.
Recently much interest has arisen
in the electroweak string,\cite{Nambu,Vach}
a type of nontopological soliton
occurring in the Weinberg-Salam model,
and in its possible implications
for astrophysics and cosmology.\cite{HHVW,Baryo}

The electroweak string is essentially a
Nielsen-Olesen cosmic string\cite{NO}
embedded in the Weinberg-Salam model.
Consider a simplified version of the Weinberg-Salam model
that includes only bosonic fields:
\be
L_{\rm boson} =
  -{1 \over 4} W_{\mu \nu}^a W^{a \mu \nu}
  -{1 \over 4} F_{\mu \nu}   F^{\mu \nu}
  + \left| D^L_\mu  \Phi \right|^2
  - \lambda \left( \Phi^\dagger \Phi - {\eta^2 \over 2} \right)^2
\label{eq:Lboson}
\ee
where $W^a_{\mu \nu}$ and $F_{\mu \nu}$
are field strength tensors for
the SU(2)$_L$ and U(1)$_Y$ gauge fields
$W^a_\mu$ and $B_\mu$ respectively,
and
$\Phi  = \left( \phi_1 \atop \phi_0 \right) $
is the complex Higgs doublet.
Gauge-covariant derivatives are given by
\be
D_\mu^L
= \del_\mu - {ig\over 2} \tau^a W^a_\mu  - {i g^\prime \over 2} Y B_\mu,
\qquad
D_\mu^R
= \del_\mu - {i g^\prime \over 2} Y B_\mu,
\label{eq:Deriv}
\ee
where $Y$ is the hypercharge of the field on which the
derivative acts.
The electroweak string, or ``Z-string,''\cite{Vach}
is the field configuration
\be
\Phi = {\eta f(\rho) \over \sqrt 2} \e^{i\phi} \left( 0 \atop 1 \right),\qquad
\left( Z^1 \atop Z^2 \right)
 = {2 v(\rho) \over \alpha \rho} \left( -\sin \phi \atop \cos \phi \right),
\label{eq:Zstring}
\ee
all other fields vanishing,
where $f(\rho)$ and $v(\rho)$
obey the Nielsen-Olesen equations\cite{NO}
\bea
 f^{\prime\prime}
+ {f^\prime \over \rho}
- (1-v)^2 {f \over \rho^2}
+ \lambda \eta^2 (1-f^2) f
& =  & 0, \nonumber \\
 v^{\prime\prime}
 - {v^\prime \over \rho}
+ {\alpha^2 \eta^2 \over 4} f^2 (1-v)
& = &  0
\label{eq:NO}
\eea
with boundary conditions
\be
f(0)=  v(0) = 0,
\qquad
f(\rho) \mathrel{\mathop{\longrightarrow}\limits_{\rho \to \infty}} 1,
\qquad
v(\rho) \mathrel{\mathop{\longrightarrow}\limits_{\rho \to \infty}} 1.
\label{eq:Boundary}
\ee
Recall that
$Z_\mu = \cos \theta_W \, W^3_\mu - \sin \theta_W \, B_\mu$
and
$\alpha = \sqrt{g^2 + {g^\prime}^2}$.

The main question is whether
the Z-string field configuration is stable.
Because the existence of electroweak strings
is due to energetic rather than topological reasons,
their stability
is dependent on the precise values of the parameters in the theory.
One possible mode of instability
is that the upper component $\phi_1$ of the Higgs field
may develop a non-zero value,
allowing the Z-string to unwind.
(There are other modes of instability as well.)
To determine whether the string is stable
to small perturbations in this direction,
one computes the change in the bosonic field energy
\be
\Delta \Eb [\phi_1] = \Eb [f, v; \phi_1] - \Eb [f, v; 0].
\label{eq:DeltaEb}
\ee
If the Z-string is stable,
$\Eb [f,v;0]$ is a local minimum of the energy functional,
and $\Delta \Eb [\phi_1]$ will be quadratic
with a positive coefficient for any perturbation $\phi_1$.
To determine whether this is so,
one inserts the ansatz
$\phi_1 = (\eta/\sqrt{2})  g(\rho) \e^{i\omega t}$
into the equations of motion to obtain
the eigenvalue equation
\be
-  g^{\prime\prime}
- {g^\prime \over \rho}
+ (\cos 2\theta_W)^2 v^2 {g \over \rho^2}
+ \lambda \eta^2 (f^2-1) g
 =   \omega^2 g.
\label{eq:eigenvalue}
\ee
If this equation has no negative $\omega^2$ eigenvalues,
the Z-string is stable under all perturbations in $\phi_1$.
Hindmarsh calculated the eigenvalues numerically for the
special case $\sin^2 \theta_W = 1$ and found
that the Z-string is only stable when the Higgs mass
is less than the Z-boson mass.\cite{Hind}
A more involved analysis\cite{JPV}
of stability under more general perturbations
revealed that in addition
the Z-string is stable only
for $\sin^2 \theta_W$ close to unity,
a region that obviously does not include the physical world.

Various authors have tried to increase
the range of stability of the Z-string
by changing the gauge or Higgs sectors of the model.\cite{HHVW,VW,Stable}
Another idea, familiar from the study of nontopological solitons,
is to add particles that gain their mass from the Higgs mechanism.
These particles remain massless at the center of the string
where the Higgs field vanishes,
and their presence at the core
would resist the string's dissolution,
because that would increase their energy.
Indeed, the presence of charged
scalar bound states was shown
to lower the value of $\sin^2 \theta_W$
for which the string is stable.\cite{VW}

It has been suggested\cite{VW} that
a similar enhancement of stability
could be attained by using fermion bound states on the Z-string.
The existence of Z-string zero modes,
fermion states localized on the string with zero energy,
lends support to this idea.\cite{EP,GV,MOQ}
Another advantage of this suggestion is
that fermions are already contained in
the standard electroweak model.
In the following sections,
we consider the effect of standard model fermions
on the stability of the Z-string.

\section{Fermion zero modes}

To discover the effect of fermion fields on the Z-string,
we must first determine the fermion spectrum
in the background of a Z-string.
The Lagrangian of the Weinberg-Salam model, including fermions, is
\be
L = L_{\rm boson} + \sum L_{\rm quark} + \sum L_{\rm lepton}.
\label{eq:Lag}
\ee
Each quark doublet contributes a term
\bea
L_{\rm quark}
& = &
  \Psibar^L i \Dslash^L \Psi^L
+ \psibar_+^R i \Dslash^R \psi_+^R
+ \psibar_-^R i \Dslash^R \psi_-^R
\nonumber\\
& &
- {G_+} \left( \Psibar^L \Phitil \psi_+^R
            + \psibar_+^R \Phitil^\dagger \Psi^L \right)
- {G_-}  \left( \Psibar^L \Phi \psi_-^R
            + \psibar_-^R \Phi^\dagger \Psi^L \right)
\label{eq:Lquark}
\eea
where $\Psi^L = \left( \psi_+^L \atop \psi_-^L \right)$,
and
$\Phitil = i \tau^2 \Phi^* = \left( \phi_0^* \atop -\phi_1^* \right) $.
Each lepton doublet contributes the same term,
absent any pieces containing $\psi_+^R$.
We neglect interfamily mixing.

The Dirac equation in the Z-string background has the form
\bea
\gamma^\mu ( i \del_\mu - {\alpha \ell_{\pm}\over 2} Z_\mu ) \ppm^L
- m_\pm f(\rho) \e^{\mp i\phi} \ppm^R
&=&  0, \nonumber\\
\gamma^\mu ( i \del_\mu - {\alpha r_{\pm}\over 2} Z_\mu ) \ppm^R
- m_\pm f(\rho) \e^{\pm i\phi} \ppm^L
&=& 0,
\label{eq:Dirac}
\eea
where $m_\pm = G_\pm \eta / \sqrt2$,
$\ell_\pm = (y \pm 1) \sin^2 \theta_W \mp 1 $,
and $   r_\pm = (y \pm 1) \sin^2 \theta_W $,
with $y$ the hypercharge of the left-handed doublet $\Psi^L$.
This equation has zero-energy modes,\cite{JRos,EP,GV,MOQ}
which obey $ \gamma_0 \gamma_3 \ppm = \pm \ppm $.
Using the chiral representation for the Dirac matrices
\be
\gamma^5 = \pmatrix{ 1 & 0  \cr 0 & -1 \cr}, \quad
\gamma^0 = \pmatrix{ 0 & -1 \cr -1 & 0 \cr}, \quad
\gamma^i = \pmatrix{ 0 & \tau^i \cr -\tau^i & 0 \cr},
\label{eq:Gamma}
\ee
and recalling that
$\psi_\pm^L = {1\over 2} (1 -\gamma^5) \psi_\pm$
and
$\psi_\pm^R = {1\over 2} (1 +\gamma^5) \psi_\pm$,
one may write the zero mode solutions as
\be
\psi_{\pm,0}  ~=~
\e^{ \pm \int_0^\rho  \left[\ell_\pm v(\rrho) / \rrho \right] \d \rrho }
\left( {i \over m_\pm f(\rho)} \P^\prime(\rho) \chi_\pm \atop
        \P (\rho) \chi_\mp \right)
\label{eq:Zeromode}
\ee
where $\chi_+ = \left( 1 \atop 0 \right)$, $\chi_- = \left( 0 \atop 1 \right)$,
with $\P(\rho)$ obeying the equation
\be
\P^{\prime\prime}
\, - \, {f^\prime \over f} \P^\prime
\, \pm \, {(\ell_\pm + r_\pm) v\over \rho } \P^\prime
\, - \,  m^2_\pm f^2 \P \, = \, 0
\label{eq:P}
\ee
and normalized by
\be
\int \d^3 x \,
\e^{ \pm 2 \int_0^\rho  \left[\ell_\pm v(\rrho) / \rrho \right] \d \rrho }
\left[ \P^2 + \left( \P^\prime/ m_\pm f \right)^2 \right] = 1
\label{eq:Norm}
\ee
For the neutrino ($m_+ = 0$, $\ell_+ = -1$),
Eq.~(\ref{eq:P}) has the simple solution $P_+ = 1$,
but by Eq.~(\ref{eq:Boundary}),
$ \psi_{+,0}
\mathrel{\mathop{\longrightarrow}\limits_{\rho \to \infty}}
1/\rho$,
so the zero mode is not normalizable,
at least for a straight infinite string
(but see ref.~11).
Eq.~(\ref{eq:P}) has the explicit solution
\be
\P (\rho) = N \e^{ - \int_0^\rho  m_\pm f(\rrho) \d \rrho }
\label{eq:Explicit}
\ee
for the special case $ y=0$ and $\cos^2 \theta_W = {1\over2}$.

The existence of zero modes generates a
$2^N$-fold degeneracy of the Z-string ground state,
where $N$ is the number of quark and charged lepton flavors.
The ground state of the string
will have the global quantum numbers of each fermion flavor,\cite{JReb}
either $\half$ or $-\half$,
depending on whether the corresponding zero mode is occupied or not.
The occupation of the zero modes will
not alter the Nielsen-Olesen equations (\ref{eq:NO}) for the string profile,
because the fermion source term for the $\phi_0$ and $Z_\phi$ fields
vanishes for the zero modes.

In the $(3+1)$-dimensional context of the Z-string,
the zero-energy solution (\ref{eq:Zeromode}) generates
a whole family of solutions of the Dirac equation
\be
\psi_{\pm,p} (\rho,z,t)
{}~=~  \e^{ ipz - i\eps_{\pm,p} t } ~\psi_{\pm,0} (\rho)
\label{eq:Massless}
\ee
with energies
\be
\eps_{\pm, p} = \pm p .
\label{eq:Dispersion}
\ee
These solutions correspond to
massless chiral fermions confined to the Z-string;
the up-type quarks run up the string (in the $+z$ direction)
and the down-type quarks and charged leptons run down the string
(in the $-z$ direction) at the speed of light.
In addition to these ``massless'' solutions of the Dirac equation,
there are many ``massive'' solutions,
whose energies are separated from zero by a finite gap.

\section{Fermion vacuum energy}

What effect do the fermion zero modes described in the previous
section have on the stability of the Z-string?
Earnshaw and Perkins\cite{EP}
pointed out that the fermion zero mode
provides a non-vanishing source term
in the equation of motion for $\phi_1$.
This violates the ``Vachaspati existence criterion''\cite{Vach}
and would appear to imply that
the Z-string configuration with $\phi_1=0$
is not an extremum of the energy.
Such a conclusion, however, would be premature.

The reason that the zero modes are a source for $\phi_1$
is that the presence of a non-zero value of $\phi_1$
lifts the degeneracy between the
$\psi_{+,0}$ and $\psi_{-,0}$ zero modes,
one linear combination of the zero modes shifting up
and the orthogonal combination shifting down.
The lower eigenstate is filled in the ground state of the Z-string,
so its descent lowers the Z-string energy.
Before drawing any conclusions about the overall stability
of the Z-string, however,
we must determine the effect of the $\phi_1$ perturbation
on the rest of the fermion eigenenergies.

The effective energy of the Z-string ground state is
\be
\Ee = \Eb + \Ef
\label{eq:Effective}
\ee
where $\Eb$ is the bosonic field energy
and $\Ef$ the fermion vacuum energy in the Z-string background
(\ie, the energy of the filled Dirac sea).
The change $\Delta \Eb [\phi_1]$
under a small perturbation $\phi_1$
was considered above (\ref{eq:DeltaEb});
the change in the fermion vacuum energy is
\be
\Delta \Ef [\phi_1]
= \sum_{ \eps_{+,n} <0}  \Delta \eps_{+,n} [\phi_1]
+ \sum_{ \eps_{-,n} <0}  \Delta \eps_{-,n} [\phi_1]
+ \delta E [\phi_1]
\label{eq:DeltaEf}
\ee
where
$ \Delta \eps_{\pm,n} [\phi_1] $
denotes the shift of the Z-string Dirac eigenenergies
$\eps_{\pm, n}$ under the perturbation,
and the sum is over negative-energy eigenvalues only.

We compute the eigenvalue shifts
$ \Delta \eps_{\pm,n} [\phi_1] $
perturbatively in $\phi_1$.
Because the perturbation is off-diagonal in the $+$ and $-$ fields,
the leading shift is second order,
and the change in fermion vacuum energy is
\bea
\Delta \Ef [\phi_1]
&=&
 \sum_{\eps_{+,n}<0} \sum_{\eps_{-,m}>0}
 {\left|  \int \d^3 x
 \left( {G_-} \psibar_{-,m}^R  \psi_{+,n}^L
- {G_+} \psibar_{-,m}^L  \psi_{+,n}^R  \right)
\phi_1^* \right|^2 \over \eps_{+,n} - \eps_{-,m} }
\nonumber\\
&+&
\sum_{\eps_{-,n}<0} \sum_{\eps_{+,m}>0}
{\left|  \int \d^3 x
  \left( {G_-} \psibar_{+,m}^L  \psi_{-,n}^R
       - {G_+} \psibar_{+,m}^R  \psi_{-,n}^L \right)
 \phi_1 \right|^2 \over \eps_{-,n} - \eps_{+,m} } + \delta E [\phi_1].
\nonumber\\
\label{eq:Second}
\eea
The sums over intermediate energy eigenstates $\eps_{\pm, m}$
include only positive-energy states;
the contributions from negative-energy intermediate states
cancel between the two sums.
The sums in Eq.~(\ref{eq:Second}) diverge in the ultraviolet.
The Z-string is not responsible for this,
for the same divergence occurs
in the usual constant field background.
In that case,
the divergence is cancelled
by adding a counterterm $\delta E [\phi_1]$.
The same counterterm will suffice to
render $\Delta \Ef [\phi_1]$ ultraviolet finite.

Let us evaluate the shifts in eigenenergies
corresponding to the massless solutions (\ref{eq:Massless}) more explicitly.
First, restrict the perturbation $\phi_1$
to a constant (complex) value $\eta g /\sqrt2 $ over the region
where the zero mode wavefunction (\ref{eq:Zeromode}) is appreciable
(but let $\phi_1 \to 0$ as $\rho \to \infty$).
Second,
noting that $\Ef$
is proportional to the length of the string (as is $\Eb$),
consider a Z-string of length $L$.
Periodic boundary conditions on the fermion wavefunctions
restrict the $z$-momenta to $ p = 2\pi n / L$.
Taking $L$ large,
the sum of the energy shifts of the massless states becomes
\be
\Delta \Efz [g]
 = -{L \over \pi} |g|^2 |A|^2 \int_0^\Lambda {\d p \over 2p}
\label{eq:IRdivergent}
\ee
with
\be
A= 2\pi iL \int \rho \d \rho
{ \left( P_+ P_- \right)^\prime \over f}
\exp \left({\int_0^\rho \left[ (\ell_+ - \ell_-) v(\rrho)
               / \rrho \right] \d \rrho}\right)
\label{eq:A}
\ee
where we have included only massless intermediate energy states.
Among this subset of intermediate states,
a selection rule ensures that the perturbation
only couples the eigenstate $\psi_{\pm,p}$
to the eigenstate $\psi_{\mp,-p}$.

The integral over momenta (\ref{eq:IRdivergent}) diverges
both in the ultraviolet and in the infrared.
As previously noted,
the ultraviolet divergence is cancelled by counterterms;
the infrared divergence signals
the breakdown of the perturbative expansion
when the energy denominator $2p$
becomes smaller than the perturbation $g$.
We redo the calculation for states with small $p$,
now treating $p$ as part of the perturbation.
The unperturbed states are now degenerate;
degenerate perturbation theory yields the perturbed energies
\be
\left| \matrix{
\eps - p  & g A   \cr  g^*  A^*  & \eps + p \cr} \right| = 0
\qquad \Rightarrow \qquad
\eps = \pm \sqrt{ p^2 + |gA|^2 }
\label{eq:Degenerate}
\ee
As mentioned above,
the degenerate zero mode ($p=0$) states
are resolved into $ \eps = \pm |gA|$.
This improved calculation yields an infrared-finite result
\bea
\Delta \Efz [g]
&=&
- {L\over 2\pi} \int_{-\Lambda}^\Lambda \d p
\left( \sqrt{ p^2 + |gA|^2 } - |p| \right)
\nonumber\\
& \mathrel{\mathop{\longrightarrow}\limits_{\Lambda \to \infty}} &
- {L \over 4 \pi} |gA|^2
\left[1+\log \left( 4\Lambda^2 \over |gA|^2 \right) \right]
\label{eq:IRfinite}
\eea
The ultraviolet divergence is absorbed by the counterterm,
leading to a completely finite expression
for the change in the fermion vacuum energy
(per unit length) under the perturbation $\phi_1$:
\be
\Delta \Ef [g]
=  L \left(  {|A|^2 \over 4 \pi} |g|^2  \log |g|^2 + \Cf |g|^2 \right)
\label{eq:DeltaEfg}
\ee
The coefficient $\Cf$ receives contributions
from the shifts of the massive Dirac eigenvalues
as well as from the finite part of the counterterm.
Each quark doublet contributes a term of the form (\ref{eq:DeltaEfg}) to
the fermion vacuum energy (with different values of $A$);
the charged leptons do not contribute
because their eigenenergies are not shifted by the perturbation
(in the absence of normalizable neutrino zero modes).

The change in the bosonic field energy
for the perturbation we are considering
has the form\cite{Vach}
\be
\Delta \Eb [g] = L \Cb |g|^2
\label{eq:DeltaEbg}
\ee
where the sign of $\Cb$,
which depends on the parameters of the model,
determines whether the bosonic Z-string (without fermions)
is stable in the direction of this perturbation.
Thus, the effective energy of the ground state of the
Z-string is
\be
\Ee [g]
= \Ee [0]
+  L \left(  C |g|^2
+ {1 \over 4 \pi}
|g|^2 \log |g|^2
\sum_{\rm quark} |A|^2
\right)
\label{eq:Effectiveg}
\ee
Observe first that
$|g| = 0$
is an extremum of this expression,
so even in the presence of fermions
the Z-string configuration (\ref{eq:Zstring}) with $|\phi_1| = 0$
remains a solution of the equations of motion.
This extremum, however,
is necessarily a maximum,
regardless of the value of $C$.
(A similar phenomenon occurs in two dimensions.\cite{CY})
Hence, the Z-string ground state is unstable
to perturbations in $\phi_1$
for {\it all} values of the parameters of the Weinberg-Salam model.

\begin{figure}[t]
\epsfysize=4.2in
\hspace*{0.0in}
\epsffile{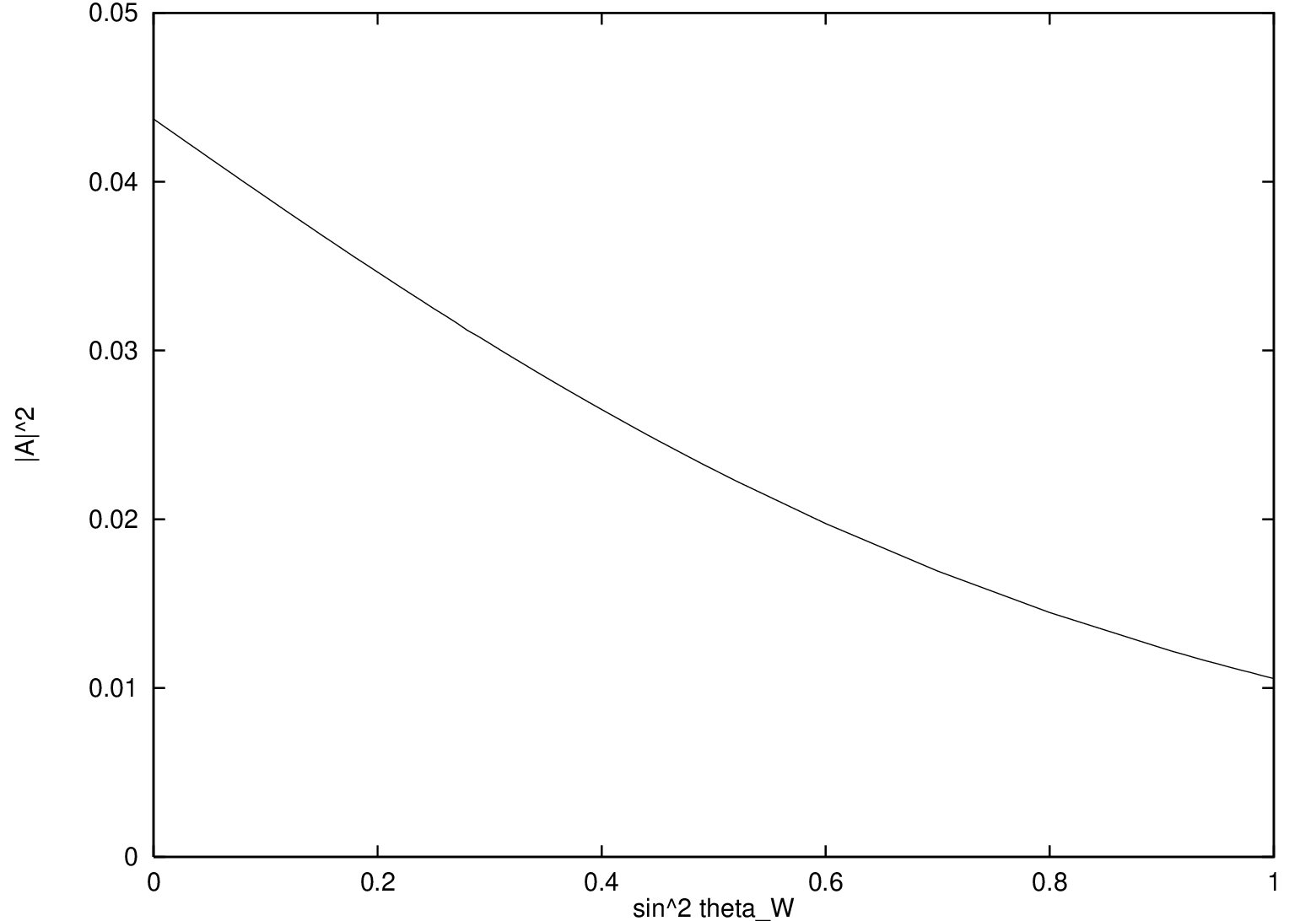}
\fcaption{The value of $|A|^2$ as a function of $\sin^2 \theta_W$.}
\end{figure}

To determine the coefficient of the
$ |g|^2 \log |g|^2 $ term in the fermion vacuum energy,
we have numerically computed the zero-mode wavefunctions (\ref{eq:Zeromode})
for the various quarks,
and used them to calculate $A$ using Eq.~(\ref{eq:A}).
The value of $|A|^2$ is a monotonically increasing
function of the quark masses,
with the third generation of quarks making
the dominant contribution to $ \sum_{\rm quark} |A|^2 $.
In Fig.~1,
we have plotted $|A|^2$
for the top-bottom quark pair
as a function of $\sin^2 \theta_W$,
taking
$m_{\rm bottom} = 5$~GeV,
$m_{\rm top} = 175$~GeV,
and, arbitrarily,
$m_{\rm Higgs} = m_Z$.
(The value of $|A|^2$ depends only weakly on the Higgs mass.)
Thus, for $\sin^2 \theta_W = 0.23$,
the value of the coefficient $ \sum_{\rm quark} |A|^2 $,
including a color factor of three,
is of order 0.1.

For a Z-string of large but finite length $L$,
the fermion energy (\ref{eq:IRfinite}) becomes a sum over momenta,
given by
\be
\Delta \Efz [g]
= - | gA |
- { L |gA|^2  \over 2 \pi }
\left[ \log \left(L \Lambda \over 2\pi\right) + \gamma \right] + \cdots
\ee
(where $\gamma = 0.577\ldots $) in the limit $ g \ll 1/L$.
The leading term linear in $g$ can be cancelled
by populating both (or neither) zero modes.
The subleading term then contributes energy per unit length
quadratic in $g$ with a negative coefficient
that diverges as $\log L$.
Thus, regardless of the bosonic contribution
(\ref{eq:DeltaEbg}),
the Z-string ground state is unstable for large $L$,
the same conclusion reached above.

\section{Conclusions and Outlook}

It has been speculated that the presence of fermions
would enhance the stability of the Z-string.
We have shown\cite{Nacu} that, on the contrary,
standard model fermions {\it destabilize} Z-strings.
More precisely,
the lowest-energy (or ground) state of the Z-string
is always a local maximum of the energy functional
with respect to (at least) one of the modes of instability.
The ground state of the Z-string is therefore unstable
for {\it all} values of the parameters
of the Weinberg-Salam model.

This instability results from the fermion vacuum energy,
which also has an important effect
on other types of solitons.\cite{Vacuum}
One cannot consistently consider
the effects of positive-energy fermion states
without also taking account of the
(filled) negative-energy states,
particularly because,
with the existence of zero modes,
there is no gap between them.
We have shown that the contribution
to the energy functional of the filled Dirac sea,
\ie, the fermion vacuum energy,
is a local maximum for the Z-string.

This does not necessarily mean that
there exists no stable nontopological string configuration.
A (locally) stable string with $\phi_1$ slightly displaced
from zero may exist, though that remains to be demonstrated.
What we are saying is that
the simple Nielsen-Olesen string embedded
into the Weinberg-Salam model
with all other fields vanishing
is necessarily unstable.

Most attempts to increase the stability of the Z-string
do so by increasing the coefficient $C$ in Eq.~(\ref{eq:Effectiveg});
this will not work here since no coefficient,
however positive,
can outweigh the negative curvature at $|g|=0$
caused by the $|g|^2 \log |g| $ term.
The only way to overcome this term
is to occupy some of the {\it positive}-energy fermion states.
This requires not just a single fermion
(as in the case of nontopological solitons)
but rather a finite density of fermions along the Z-string.
If the string holds $\zeta$ positive-energy fermions
per unit length of types $+$ and $-$,
the effective energy will change by
\bea
\Delta E [g]
 & =&
 {L\over 2\pi} \int_{-2\pi\zeta}^{2\pi\zeta} \d p
\left( \sqrt{ p^2 + |gA|^2 } - |p| \right)
\nonumber\\
& \mathrel{\mathop{\longrightarrow}\limits_{ |gA| \ll 2\pi\zeta}}&
 {L \over 4 \pi} |gA|^2
\left[1+\log \left( 16 \pi^2 \zeta^2 \over |gA|^2 \right) \right]
\label{eq:Finitedensity}
\eea
If the Z-string carries
a non-zero density of quarks of {\it each} flavor,
the change in energy
will cancel the $ |g|^2  \log |g|^2 $ piece in Eq.~(\ref{eq:Effectiveg}),
rendering the total energy proportional to $|g|^2$.
It is therefore possible that a higher-energy state of the Z-string,
with some finite quark density,
could be stable.

\section{Acknowledgements}

The work of S.~K.~was supported by a
Surdna Foundation Undergraduate Research Fellowship.
S.~N.~wishes to thank T.~Vachaspati for useful discussions.

\section{References}

\vspace{-0.25cm}

\def\PLB{ \it  Phys. Lett.         \bf B}
\def\NPB{ \it  Nucl. Phys.         \bf B}
\def\PRD{ \it Phys. Rev.          \bf D}
\def\PRL{ \it Phys. Rev. Lett.    \bf  }

\end{document}